\documentclass[preprint,showpacs,preprintnumbers,amsmath,amssymb,superscriptaddress,nofootinbib]{revtex4}

\begin{document}


\title{Topological Zero-Thickness Cosmic Strings}


\author{Yi-Shi Duan}

\author{Zhen-Bin Cao}
\email{caozhb04@lzu.cn}

\affiliation{Institute of Theoretical Physics, Lanzhou University,
Lanzhou, China 730000}

\date{\today}


\begin{abstract}

In this paper, based on the gauge potential decomposition and the
$\phi-$mapping theories, we study the topological structures and
properties of the cosmic strings that arise in the Abelian-Higgs
gauge theory in the zero-thickness limit. After a detailed
discussion, we conclude that the topological tensor current
introduced in our model is a better and more basic starting point
than the generally used Nambu-Goto effective action for studying
cosmic strings.

\end{abstract}

\maketitle

\smallskip

\section{Introduction}                                                          

Cosmic strings, the idea of which was very popular in the 1980's and
much of 90's, are linear topological defects that may be formed at
phase transitions in the early universe
\cite{cos-string,book-cos-string}. Though the recent observations of
the cosmic microwave background by WMAP \cite{WMAP} showed beyond
doubt that cosmic strings could not provide an adequate explanation
for the primordial density fluctuations, but could only contribute
at the level of a few percent, they still have very fruitful
implications for cosmology, for instance, they can generate
gravitational lensing effects, gravitational waves, and so on.
Meanwhile, the recent developments in superstring theory suggest
that cosmic superstrings (macroscopic fundamental strings or
one-dimensional Dirichlet branes) can play a role very similar to
that of cosmic strings. Also, on the observational side, though the
galaxy image pair CSL-1 \cite{CSL-1,CSL-1-2} has been excluded as a
candidate for a cosmic string lens \cite{exclud-CSL-1}, it now still
gives us the impression that, cosmic strings might provide a best
observational window to the very early universe, the extremely high
energy physics, and possibly to superstring theory, and thus have
been of particular interest recently.

Though the cosmological evolution of the cosmic strings has been
studied extensively, they were mostly focused on their dynamical
properties. We note that, their topological properties are also very
important and worth investigating. In this paper, by making use of
the gauge potential decomposition and the $\phi-$mapping theories
\cite{decomposition-A,decomposition-A-2,phi-mapping,phi-mapping-2},
we study the topological properties of the cosmic strings that arise
in the Abelian-Higgs model with a broken U(1) gauge symmetry in the
zero-thickness limit in detail. Since almost all of the ideas can
carry over to cosmic superstrings, our results are also useful in
superstring theory.

The paper is organized as follows: In Sec.
\ref{formationandstructure}, we give a slightly detailed review of
the topological structures of the cosmic strings, showing that
they can be totally decided by the distribution of some complex
scalar field. Also we show that as we can deduce the Nambu-Goto
effective action of the strings, all the cosmic strings in our
model have a zero thickness. In Sec.
\ref{property-cosmic-strings}, we study the topological properties
of the zero-thickness cosmic strings, showing that cosmic strings
not only represent trapped energy, but also represent trapped
flux. We point out that, the winding numbers of the strings are
the products of their Hopf indices and Brouwer degrees, and the
flux carried by a string is its winding multiple of the flux
quantum $\frac{2\pi}{e}$. For application, we show a nature
picture of the evolution of the early universe. The conclusion is
given in Sec. \ref{conclusion}, in which we suggest a better and
more basic starting point for studying cosmic strings than the
generally used Nambu-Goto effective action.


\section{Topological Structure of Cosmic Strings}                               
\label{formationandstructure}

There are various kinds of cosmic strings in the literature
\cite{cos-string,book-cos-string}, in this paper, we mainly focus
our attention on the cosmic strings that generally appear in the
Abelian-Higgs model with a broken U(1) gauge symmetry, though we
note that our results can be generalized to other cases easily.
The Lagrangian density is given by
\begin{equation}\label{lagrangian}                                              
{\cal L}=-\frac{1}{4}F_{\mu\nu}F^{\mu\nu}+(D_\mu\phi)^*D^\mu\phi
  -V(\phi),
\end{equation}
where $\phi(x)$ is a complex scalar field, which can also be thought
of as a pair of real fields $\phi^{1,2}$, with
$\phi=\phi^1+i\phi^2$, $A_\mu$ is a gauge field,
$F_{\mu\nu}=\!\partial_\mu A_\nu\!-\!\partial_\nu A_\mu$ is the
corresponding field strength tensor, and the covariant derivative
$D_\mu\!=\!\partial_\mu-ieA_\mu$. Usually the potential $V(\phi)$ is
a function only of $|\phi|$, and here is taken to be the form
\begin{equation}\label{potential}
V(\phi)=\frac{\lambda}{4}(\phi^*\phi-\eta^2)^2,
\end{equation}
where $\eta$ is the vacuum expectation value (VEV) of $|\phi|$.
Also there are two dimensionless parameters, the gauge coupling
constant $e$ and the Higgs self-coupling $\lambda$. Meanwhile, the
metric is taken to be that of the spatially flat expanding
universe
\begin{equation}
ds^2=g_{\mu\nu}dx^\mu dx^\nu=-dt^2+a^2(t)(dx^2+dy^2+dz^2).
\end{equation}

For cosmic string formation, there are several mechanisms, such as
the Kibble mechanism, the thermal fluctuations of the magnetic
field, etc. Here, we assume the cosmic strings have already
formed, and discuss their topological structures.


As the gauge potential $A_\mu$ is initially introduced in the
covariant derivative of $\phi(x)$, it is natural to think that there
is some relationship between the distributions of $A_\mu$ and
$\phi(x)$. According to the gauge potential decomposition theory
\cite{decomposition-A,decomposition-A-2} proposed by one of the
authors (Duan), it can be seen clearly that, besides a gauge
transformation term, $A_\mu$ is totally decided by $\phi(x)$
\begin{equation}\label{gaugepotential}                                          
A_\mu=\frac{1}{e}\epsilon_{ab}\frac{\phi^a}{\|\phi\|}
      \partial_\mu\frac{\phi^b}{\|\phi\|}
      +\frac{1}{e}\partial_\mu\theta,
\end{equation}
where $\|\phi\|\!=\!\sqrt{\phi^a\phi^a}$ and $\theta$ is only a
phase factor.

In order to study the topological structures of the strings, we
define a second order topological tensor current as follow
\begin{equation}\label{def-j}                                                   
j^{\mu\nu}=\frac{1}{2\sqrt{-g}}\epsilon^{\mu\nu\lambda\rho}
           F_{\lambda\rho}
          =\frac{1}{2\sqrt{-g}}\epsilon^{\mu\nu\lambda\rho}
           (\partial_\lambda A_\rho-\partial_\rho A_\lambda),
\end{equation}
where $g$ is the determinant of the metric $g\!=\!det(g_{\mu\nu})$.
Substituting (\ref{gaugepotential}) into (\ref{def-j}), it yields
\begin{equation}\label{decom-form-j}
j^{\mu\nu}=\frac{1}{e\sqrt{-g}}\epsilon^{\mu\nu\lambda\rho}
           \epsilon_{ab}\partial_\lambda\frac{\phi^a}{\|\phi\|}
           \partial_\rho\frac{\phi^b}{\|\phi\|},
\end{equation}
which shows that $j^{\mu\nu}$ is antisymmetric and identically
conserved
\begin{equation}\label{conserve-j}
\nabla_\mu j^{\mu\nu}
  =\frac{1}{\sqrt{-g}}\partial_\mu(\sqrt{-g}j^{\mu\nu})
  =0.
\end{equation}
Following the $\phi-$mapping theory \cite{phi-mapping,phi-mapping-2}
which was also proposed by author Duan, this topological tensor
current can be expressed in a compact $\delta-$function form
\begin{equation}\label{j-delta}                                                 
j^{\mu\nu}=\frac{2\pi}{e}\delta(\phi)J^{\mu\nu}
           (\frac{\phi}{x}),
\end{equation}
where $J^{\mu\nu}(\frac{\phi}{x})$ is the general Jacobian tensor
defined as
\begin{equation}
\epsilon^{ab}J^{\mu\nu}(\frac{\phi}{x})=
             \frac{\epsilon^{\mu\nu\lambda\rho}}{\sqrt{-g}}
             \partial_\lambda\phi^a\partial_\rho\phi^b.
\end{equation}
From (\ref{j-delta}) one sees clearly that $j^{\mu\nu}$ is
nontrivial only when $\phi\!=\!0$, or equivalently
\begin{equation}\label{phi=zero}
\left \{
\begin{array}{c}
\phi^1(x)=0\\   \phi^2(x)=0\\
\end{array}.
\right.
\end{equation}

Suppose that for the equations (\ref{phi=zero}), there are $K$
different solutions. According to the implicit function theorem,
when the regular condition of $\phi(x)$ at which the rank of the
Jacobian matrix $(\partial_\mu\phi^a)$ is 2 satisfies, these
solutions can be expressed as
\begin{equation}                                                                
x^\mu=x^\mu_i(u^1,u^2),  \;\;i=1,\cdots,K,
\end{equation}
where the subscript $i$ represents the $i$th solution which is a
two dimensional submanifold spanned by the parameters
$u^c(c\!=\!1,2)$ with the metric tensor
$\gamma_{cd}\!=\!g_{\mu\nu}(\partial x^\mu/\partial u^c)(\partial
x^\nu/\partial u^d)$ and called the $i$th singular submanifold
$N_i$ in the spacetime. For each $N_i$ it can be proved that there
exists a local two dimensional normal submanifold $M_i$ in the
spacetime spanned by the parameters $v^A(A\!=\!1,2)$ with the
metric tensor $g_{AB}\!=\!g_{\mu\nu}(\partial x^\mu/\partial
v^A)(\partial x^\nu/\partial v^B)$, which is transversal to $N_i$
at the intersection point $p_i$. Then at the regular point $p_i$,
the regular condition can be expressed explicitly as
\begin{equation}\label{regularcondition}
J(\frac{\phi}{v})=\frac{\partial(\phi^1,\phi^2)}{\partial(v^1,v^2)}
\neq0.
\end{equation}

By using the $\delta-$function theory, one can prove that
\begin{equation}\label{delta-phi}
\delta(\phi)=\sum_i\frac{\beta_i\zeta_i}{J(\frac{\phi}{v})|_{p_i}}
             \delta(N_i),
\end{equation}
where $\beta_i$ is a positive integer called the Hopf index of
$\phi-$mapping and $\zeta_i\!=\!signJ(\phi/v)|_{p_i}\!=\!\pm1$ is
the Brouwer degree. $\delta(N_i)$ is the $\delta-$function on the
singular submanifold $N_i$ with the expression
\begin{equation}\label{delta-N}
\delta(N_i)=\!\int\!_{N_i}\delta(x-x_i(u^1,u^2))\sqrt{-\gamma}d^2u.
\end{equation}
Substituting (\ref{delta-phi}) into (\ref{j-delta}), one gets the
final expansion form of $j^{\mu\nu}$ on the $K$ singular
submanifolds
\begin{equation}\label{final-j}                                                 
j^{\mu\nu}=\frac{2\pi}{e}\sum_{i}\beta_i\zeta_i
           \delta(N_i)\frac{J^{\mu\nu}(\frac{\phi}{x})}
           {J(\frac{\phi}{v})|_{p_i}},
\end{equation}
or, in terms of parameters $y^A\!=\!(v^1,v^2,u^1,u^2)$,
\begin{equation}\label{final-j2}
j^{AB}=\frac{2\pi}{e}\sum_{i}\beta_i\zeta_i
           \delta(N_i)\frac{J^{AB}(\frac{\phi}{y})}
           {J(\frac{\phi}{v})|_{p_i}}.
\end{equation}

Now, we see that, if taking $u^1$ and $u^2$ to be timelike
evolution parameter $t$ and spacelike string parameter $\sigma$
$(\text{i.e.}\; u^1\!=\!t,u^2\!=\!\sigma)$, respectively, as it
can be proved that
$J^{\mu\nu}(\frac{\phi}{x})/J(\frac{\phi}{v})|_{p_i}$ or
$J^{AB}(\frac{\phi}{y})/J(\frac{\phi}{v})|_{p_i}$ has the
dimension of velocity, the inner topological structures of
$j^{\mu\nu}$ or $j^{AB}$ just represent $K$ isolated singular
strings moving in the universe. These singular strings are just
the cosmic strings, and the two dimensional singular submanifolds
$N_i(i\!=\!1,\cdots,K)$ are their world sheets. Meanwhile, we can
classify these cosmic strings in terms of their Brouwer degrees
$\zeta_i$: a string is called a cosmic string if its $\zeta>0$ and
an anti cosmic string if its $\zeta<0$.

Further, if we define the Lagrangian density of the cosmic strings
as the generalization of Nielsen's Lagrangian
\cite{nielsen,nielsen-2}
\begin{equation}                                                                
{\cal L}_{str}=-\mu\sqrt{\frac{1}{2}
g_{\mu\lambda}g_{\nu\rho}j^{\mu\nu}j^{\lambda\rho}}
=-\mu\frac{2\pi}{e}\delta(\phi)J(\frac{\phi}{v}),
\end{equation}
where $\mu$ is the string tension, which is defined as the energy
per unit length, then by using (\ref{delta-phi}) and (\ref{delta-N})
one can get the action of the strings
\begin{eqnarray}\label{string-action}
S_{str}\!&\equiv&\!\!\int\!d^4x\sqrt{-g}{\cal L}_{str}
         =-\mu\!\int\!d^4x\sqrt{-g}\frac{2\pi}{e}\delta(\phi)
          J(\frac{\phi}{v})       \nonumber \\
   \!&=&\!\frac{2\pi}{e}\sum_i\beta_i\zeta_iS_i,
\end{eqnarray}
where
\begin{equation}\label{nambu-action-string}                                     
S_i=-\mu\!\int_{N_i}\!\sqrt{-\gamma}d^2u
\end{equation}
is exactly the Nambu-Goto effective action of the $i$th string,
which shows that the thickness of the string is zero. So that the
cosmic strings in our model all have a \emph{zero thickness}.
Actually, we can get an understanding of this in eq.
(\ref{decom-form-j}) topologically, which shows that the field
$\phi(x)$ is renormalized to a unit vector field
$n(x)=\frac{\phi(x)}{\|\phi\|}$ except at its zero points. Though
the real cosmic strings if there existed should not be zero
thickness, we can still get many important properties of them by
using this zero-thickness model, especially the topological
properties which have no relations with their thickness. And many
of the discussions in the literature are also based on the action
(\ref{nambu-action-string}) and by adding further correction
terms.

\section{Properties of Cosmic Strings}\label{property-cosmic-strings}

Now we study the topological properties of the cosmic strings.
Firstly, according to (\ref{phi=zero}), we see that, from the
mathematic point of view, the cosmic strings are just the
intersection lines of two infinite sheets, so that topologically
there are only two types of cosmic strings: the closed loops and the
infinite long strings.

Secondly, from the definition of $j^{\mu\nu}$ (\ref{def-j}), one
knows that it's the Hodge$\ast$ dual tensor of $F_{\mu\nu}$.
According to (\ref{j-delta}), in the spacetime where
$\phi(x)\!\neq\!0$,
\begin{equation}
j^{\mu\nu}=0\;\longrightarrow\;F_{\mu\nu}=0.
\end{equation}
We can understand this in two ways, one is that $A_\mu$ is a pure
gauge which has no independent physical meanings and can be chosen
to be zero, and the other is that the coupling constant
$e\!\rightarrow\!0$. Both these two cases lead to the same result
that the symmetry of $\phi(x)$ is global. This holds true before
the symmetry-breaking phase transition when all the values of
$\phi(x)$ are large. At that time, the Lagrangian density of
$\phi(x)$ corresponds to
\begin{equation}
{\cal L}_\phi=\partial_\mu\phi^*\partial^\mu\phi
              -V(\phi).
\end{equation}
This is an interesting result, for the Lagrangian has the same form
as that of inflation. Though in most inflationary scenarios, the
inflaton field is a scalar one, our complex scalar field can also
drive inflation very well. Actually, the inflationary model driven
by $L$ scalar fields $\phi_i(i\!=\!1,\cdots,L)$ has also been
discussed in \cite{multifield-inflation}.

After the symmetry is broken, $\phi(x)$ degenerates to its VEV,
during which, according to the Kibble mechanism and other
mechanisms, cosmic strings form, at the core of which $\phi(x)$ is
zero and the symmetry can be regarded as unbroken. If there are some
physical `magnetic fields' initially in the universe which cannot
exist in the broken phase, they must get confined into the cosmic
strings. This suggests that $A_\mu$ gets physical meanings or the
`magnetic fields' get coupled to $\phi(x)(e\!\neq\!0)$, just as
\cite{Rajantie03,Rajantie03-2} stated, ``the known cosmological
phase transitions are not simple symmetry breaking transitions, but
they involve a breakdown of a local gauge symmetry.'' And from this,
we see clearly that cosmic strings represent trapped `magnetic'
flux. From (\ref{final-j}) or (\ref{final-j2}), it's easy to see
that the flux carried by the $i$th cosmic string is
\begin{equation}\label{string-flux}
\Phi_i=\Phi_0W_i,
\end{equation}
where $\Phi_0\!=\!2\pi/e$ is the flux quantum, and $W_i$ is the
winding number of the $i$th string, determined by
\begin{equation}
W_i=\beta_i\zeta_i,
\end{equation}
($i$ is not summed). This is also an important result. We have to
mention that, although the Kibble mechanism only gives a
contribution of $\pm1$ to the winding number of the cosmic strings,
some other mechanisms can indeed form strings with high windings, as
discussed in \cite{Rajantie03,Rajantie03-2,high-winding}.


Now, combining with the inflationary scenarios and GUTs, we could
construct a brief picture of the evolution of the early universe:
After the big bang, the universe was in a very hot and dense state
filled with a hot `soup' of some kinds of scalar fields and maybe
some other massless particles. As the universe temperature T is very
high, $\phi(x)$ has a very large value, $|\phi|^2\!\gg\eta^2$, so
the center hump term $\eta^2$ in the potential $V(\phi)$
(\ref{potential}) is unimportant and can be
ignored\footnote{Generally, $\eta^2$ is a function of the
temperature $T$, but here we take it to be a constant for
simplicity, as detailed in \cite{kibble05}.}. This shows that the
symmetry of the field $\phi(x)$ is complete and global, and
fluctuations in any directions are equally likely. Then as the
universe expands and cools, $\phi(x)$ rolls down. If the rolling of
$\phi(x)$ is very slow, which satisfies the slow-roll conditions of
inflation, the universe is in the stage of inflation
\cite{linde83,linde83-2}, namely the expansion of the space
continues in a quasi-exponential way ($\sim\!e^{Ht}$) for a period
of time ($\triangle t\!\!\sim\!\!70H^{-1}$). Then it comes to the
final phase of inflation. The term $\eta^2$ in the potential
(\ref{potential}) becomes important, and the fluctuations of
$\phi(x)$ over the hump are not permitted. Thus the field tends to
settle towards one of its VEV $\phi\!=\!\eta e^{i\alpha}$ and couple
to the gauge field, during which, the Kibble mechanism and other
mechanisms work. Therefore the symmetry of $\phi(x)$ breaks down and
the cosmic strings form, whose structures are described by
(\ref{final-j}) or (\ref{final-j2}) and which eventually form a
random tangled network. As Jeannerot \emph{et al} \cite{jeannerot03}
showed in an interesting recent study that topological or embedded
cosmic strings formed at the end of inflation seem almost
unavoidable, here we give quite a nature way to achieve this in our
topological zero-thickness cosmic string model. And one thing that
is worth noting is that, according to (\ref{conserve-j}), as there
is no strings originally, after their formation, the total windings
of all the strings must be zero.

As an aside, from (\ref{potential}) and (\ref{j-delta}), it's
obvious that, in the core of the strings, $\phi(x)$ vanishes, but
the potential does not. Thus we see that, cosmic strings not only
represent trapped flux, but also represent trapped potential
energy (as well as gradient energy) \cite{kibble05}, (actually we
can also see this from (\ref{string-action})). And the important
thing is that, the density of this trapped energy which acts as a
remnant of the earlier high-temperature universe may be similar to
what it was before the symmetry-breaking phase transition, which
suggests that cosmic strings might provide a best observational
window to the very early universe, and to the extremely high
energy physics. Meanwhile, to say the density of the string
energy, we note that though for the zero-thickness cosmic strings,
all their cores reach zero represented by a delta function
$\delta(\phi)$, the actual cosmic strings can in general have more
than one thickness scale: the thickness of the field energy core
and the thickness of the gauge core.

\section{Conclusion}\label{conclusion}                                          

In this paper, we mainly studied the topological structures and
properties of the cosmic strings that appear in the Abelian-Higgs
model with a broken U(1) gauge symmetry in the limit of
zero-thickness. By using the gauge potential decomposition and the
$\phi-$mapping theories, and discussing the properties of the zero
points of the complex scalar field, we obtained the topological
structures of these strings, showing that under the regular
condition $J(\phi/v)\!\neq\!0$, the cosmic strings are isolated
ones moving in the universe, their winding numbers are determined
by the products of their Hopf indices and Brouwer degrees, they
represent not only trapped energy but also trapped flux and the
flux they carried are their windings multiple of the flux quantum.
Meanwhile, as we have got the Nambu-Goto effective action
(\ref{nambu-action-string}) of the cosmic strings from their
topological structures (\ref{final-j}) or (\ref{final-j2}),
varying it with respect to $x^\mu(u^c)$ will give the
Euler-Lagrange equations of the strings
\begin{equation}
x^{\mu;c}_{,c}+
\Gamma^\mu_{\nu\lambda}\gamma^{cd}x^\nu_{,c}x^\lambda_{,d}=0,
\end{equation}
where the coma and the semicolon represent the partial derivative
and the covariant derivative, respectively,
$\Gamma^\mu_{\nu\lambda}$ is the four-dimensional Christoffel
symbol, and the covariant Laplacian is
\begin{equation}
x^{\mu;c}_{,c}=\frac{1}{\sqrt{-\gamma}}
\partial_c(\sqrt{-\gamma}\gamma^{cd}x^\mu_{,d}).
\end{equation}
This suggests that though we mainly discussed the topological
properties of the cosmic strings based on their topological
structures (\ref{final-j}) or (\ref{final-j2}), these structures
can also be used to study their dynamical properties. Therefore we
conclude that the topological current (\ref{final-j}) or
(\ref{final-j2}) or the original definition form (\ref{def-j}) is
very important for the cosmic strings, for it almost describes all
their properties, and so that it can be used as a better and more
basic starting point than the generally used effective action
(\ref{nambu-action-string}) for the studying of the cosmic
strings.

Finally, we see that all the above discussions are based on the
regular condition (\ref{regularcondition}), what will happen if this
condition fails? Also, as cosmic superstrings or branes are in
essence defects of various dimensions (though not necessarily
topological), can these obtained topological properties be
generalized to them? How to do it? They are our further works.

\noindent {\bf Acknowledgments}\,                                               
One of the authors ZBC is indebted to Dr. Y. X. Liu for his much
help. This work was supported by the National Natural Science
Foundation and the Doctor Education Fund of Educational Department
of the People’s Republic of China.


\end{document}